\documentclass[aps,prl,reprint,groupedaddress,showpacs]{revtex4-1}

\usepackage{latexsym,amssymb,amsfonts,mathrsfs,color,graphicx,amsmath}

\begin{document}

\title{Hydrodynamics determines collective motion and phase behavior of active colloids}
\author{Andreas Z\"{o}ttl}
\author{Holger Stark}
\date{\today}

\affiliation{Institut f\"{u}r Theoretische Physik, Technische Universit\"{a}t Berlin, Hardenbergstrasse 36, 10623 Berlin, Germany}

\begin{abstract}
We study the collective motion of confined spherical microswimmers such as active colloids which we
model by so-called squirmers. To simulate hydrodynamic flow fields including thermal noise, we use the 
method of multi-particle collision dynamics. We demonstrate that hydrodynamic near fields acting between 
squirmers as well as squirmers and bounding walls crucially determine their collective motion. In particular, 
with increasing density we observe a clear phase separation into a gas-like and cluster phase for neutral 
squirmers whereas strong pushers and pullers more gradually approach the hexagonal cluster state.
\end{abstract}
\maketitle

The collective motion of microorganisms and artificial micro- and nanoswimmers has attracted a lot of attention among 
physicists \cite{Marchetti12,Saintillan13,Kapral13}. 
Since swimmers propel themselves autonomously through a fluid, 
they are constantly out of equilibrium and
understanding their collective properties has become a paradigm of statistical mechanics.
Simple model systems to experimentally study nonequilibrium collective motion 
are spherical colloids such as active Janus particles \cite{Palacci10,Theurkauff12,Palacci13,Buttinoni13},
active emulsion droplets \cite{Thutupalli11}, or Volvox algae \cite{Drescher09,Goldstein13}.
All these experiments have been performed in a quasi-2D geometry where the spherical particles
move (almost) in a plane bounded by one or two walls.
They show interesting nonequilibrium features like dynamic clustering \cite{Theurkauff12,Palacci13}
and swarming \cite{Thutupalli11} but also phase separation \cite{Buttinoni13}.

While equilibrium phase separation is commonly induced by attractive interparticle forces,
motility-induced phase separation simply occurs due to the activity of the particles
even without any aligning mechanisms or attractive forces.
Recently, phase separation of spherical swimmers has been studied extensively by means of
2D Brownian dynamics simulations of active Brownian disks \cite{Fily12,Redner13,Bialke13,Palacci13,Buttinoni13,Redner13b} or dumbbells \cite{Schwarz12},
and by continuum models introducing density-dependent velocities for active particles \cite{Tailleur08,Thompson11,Cates13,Stenhammer13}.
Nevertheless, the fact that active colloids typically move in an aqueous environment at low Reynolds number,
where they interact with each other and also with bounding walls via hydrodynamic flow fields, 
has not been considered in these studies.

Biological microswimmers use a non-reciprocal deformation of their cell body 
or appendages like flagella and cilia to propel themselves through a fluid \cite{Purcell77}.
Active colloids and droplets rather create a tangential slip velocity close to their surface which pushes them forward.
Hence, they can be modeled by the so-called \textit{squirmer} that propels itself by a prescribed axisymmetric 
surface velocity field \cite{Squirmer,Stone96,Ishikawa06}.
Experiments and theory show that the flow field of an active droplet 
is well approximated by the flow field of a squirmer \cite{Thutupalli11,Schmitt13}.

Several studies of the collective dynamics of squirmers in bulk exist \cite{Ishikawa07,Ishikawa08a,Evans11,Alarcon13,Molina13}.
While phase separation has not been observed, squirmers can exhibit polar order \cite{Ishikawa08a,Evans11,Alarcon13}
and collective motion in a monolayer \cite{Ishikawa08b}.
However, the monolayer is unstable and thus does not describe
a real system of confined active colloids. Recent studies on 2D squirming disks do not show cooperative
behavior \cite{Fielding12,Aguillon12} and it is argued that phase separation is suppressed \cite{Fielding12}.

Motivated by experimental systems \cite{Thutupalli11,Goldstein13}, we present here a detailed numerical 
and thereby realistic study of the collective dynamics of spherical microswimmers
in a quasi-2D geometry including their full 3D rotation.
By means of the method of multi-particle collision dynamics (MPCD),
we simulate the  hydrodynamic flow field initiated by squirmers including thermal noise.
We demonstrate that hydrodynamic near fields acting between squirmers as well as squirmers
and bounding walls crucially determine their collective motion. In particular, phase separation
into a gas-like and cluster phase depends on the squirmer type.

\begin{figure}
\includegraphics[width=.9\columnwidth]{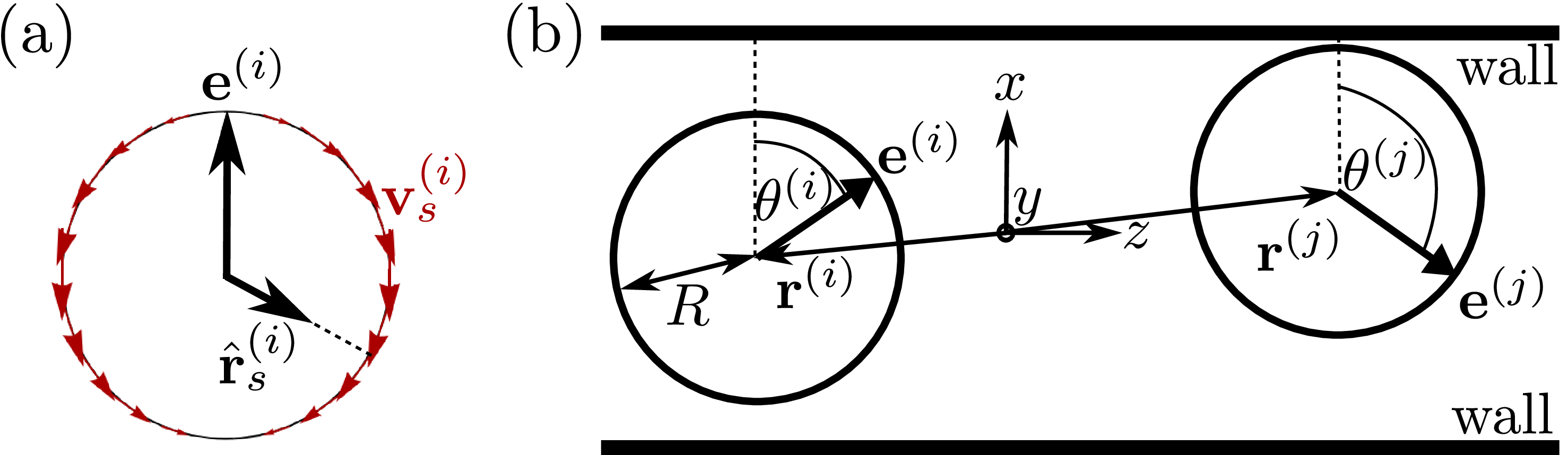}
\caption{
(a) Sketch of a spherical squirmer. The prescribed surface velocity field $\mathbf{v}_s^{(i)}$ generates a swimming velocity 
along the unit vector $\mathbf{e}^{(i)}$.
(b) Squirmers at positions $\mathbf{r}^{(i)}$ and with orientations $\mathbf{e}^{(i)}$ move in a quasi-2D geometry
bounded by two parallel planar walls.
}
\label{Fig:geometry}
\end{figure}

Our system consists of $N$ squirmers of radius $R$ that propel themselves in a fluid
by a prescribed surface velocity. For the $i$-th squirmer it is given by \cite{Ishikawa06}
\begin{equation}
\mathbf{v}_{s}^{(i)} =  B_1 \left( 1 + \beta \mathbf{e}^{(i)}\cdot \hat{\mathbf{r}}_{s}^{(i)} \right)
  \left[ (\mathbf{e}^{(i)}\cdot \hat{\mathbf{r}}_{s}^{(i)} ) \hat{\mathbf{r}}_{s}^{(i)} - \mathbf{e}^{(i)}  \right]  ,
\label{Eq:1}
\end{equation}
where $\mathbf{e}^{(i)}$ is the swimming direction and $\hat{\mathbf{r}}_{s}^{(i)}$ the
unit vector which points from the center $\mathbf{r}^{(i)}$ of the squirmer to its surface [Fig.~\ref{Fig:geometry}(a)]. 
The constant $B_1$ determines the bulk swimming speed $v_0=2B_1/3$ \cite{Squirmer}
and therefore the characteristic time scale $R/v_0$ of the system.
For $\beta<0$ the swimmer is called a \textit{pusher}, for  $\beta>0$ a \textit{puller}, 
and for $\beta=0$ a \textit{neutral squirmer}.
The names are connected to the different far fields of the squirmers,
however, when  the concentration of the squirmers is high, hydrodynamic interactions between them 
are governed by near fields initiated by the surface velocity field  of Eq.~(\ref{Eq:1}) \cite{Ishikawa06}.

The swimmers are bounded by two hard walls located at $x=\pm (1+\delta)R$ [Fig.~\ref{Fig:geometry}(b)].
We choose here a strong confinement, $\delta = 1/3$, so that the swimmer trajectories take place in 
quasi-2D but the squirmers can freely rotate in three dimensions.

To simulate the flow fields created by the squirmers we use the method of  multi-particle collision dynamics (MPCD).
The motion of a squirmer in bulk \cite{Downton09}, in confinement and under 
flow \cite{Zoettl12}, and the pairwise interaction among squirmers \cite{Goetze10}  
has already successfully been studied with MPCD. The fluid is modeled by point-like effective 
fluid particles of mass $m_0$ at a temperature $T_0$.
They perform alternating \textit{streaming} and \textit{collision} steps which are sufficient to
reproduce a flow field that solves the Navier-Stokes equations \cite{Malevanets99,Padding06,Kapral08,Gompper08}. 
In the streaming step the particles move ballistically for a time intervall $\Delta t$. 
They interact with the squirmers and the bounding walls via hard-core collisions
where the no-slip boundary condition is implemented and momentum and angular momentum are transferred.
Then, in the collision step the particles are sorted into cubic cells with edge length $a_0$.
They interact via the collision rule MPC-AT+a \cite{Gompper08} where also
virtual particles in the squirmers and the walls are included to improve the no-slip boundary condition.
This  scheme accurately reproduces the hydrodynamic flow field of a squirmer \cite{Downton09,Goetze10}
and near-field lubrication forces between squirmers \cite{Goetze10}.
To be concrete, the average number of particles per cell is $N_c=10$ and we use the
time step $\Delta t=0.02 a_0 \sqrt{m_0/k_BT_0}$ and fluid viscosity $\eta=16.05\sqrt{m_0 k_BT_0}/a_0^2$.

We study the dynamics of $N=208$ squirmers of radius $R=3$ and bulk swimming velocity $v_0 = 0.067$
at different areal densities $\phi \in [0.10,0.83]$ for $\beta \in [-3,3]$.
We use periodic boundary conditions in $y$ and $z$ direction and our confinement parameter $\delta=1/3$
is comparable, for example, to recent experiments \cite{Buttinoni13}.
Characteristic parameters of our system are the Reynolds number $\mathrm{Re}=Rv_0 \rho/ \eta = 0.12$,
the Peclet number $\mathrm{Pe}=2R v_0/D^0\approx 360$, and the persistence number 
$\mathrm{Pe}_r = v_0/(2R D_r^0) \approx 110$ \cite{Taktikos12} (again comparable to the active colloids 
in \cite{Buttinoni13}), where $D^0$ and $D_r^0$ are the respective thermal translational and rotational diffusivities,
which we measured in bulk.
The persistence number $\mathrm{Pe}_r$ compares the orientational correlation time $(D_r^0)^{-1}$ to the time to swim a distance $2R$.

\begin{figure}
\includegraphics[width=0.96\columnwidth]{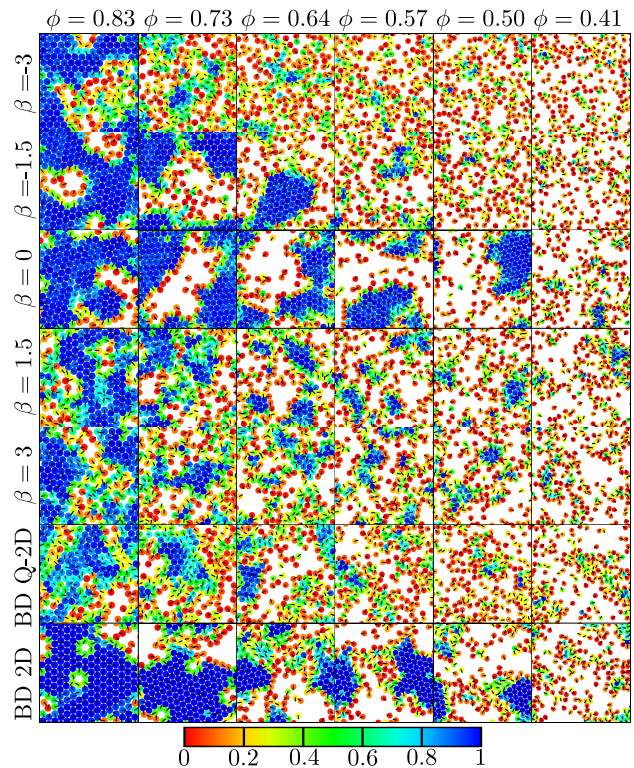}
\caption{
Typical snapshots of the collective motion of squirmers in a quasi-2D geometry depending on the 
area fraction $\phi$ and the squirmer type ($\beta$). 
Also shown are active Brownian spheres moving in quasi-2D (BD Q-2D) and active Brownian disks moving in 2D (BD 2D).
The view is from the top and the colors indicate the local bond-orientational order $|q_6|^2\in [0,1]$.
}
\label{Fig:snapshots}
\end{figure}

For each parameter set $(\beta,\phi)$ we performed eight simulation runs to have sufficient data for averaging.
After a transient the system always reached a non-equilibrium steady state which we confirmed by measuring 
the time-evolution of several order parameters which all approached a constant accompanied by fluctuations.

Typical simulation snapshots of the system in the steady state at medium and high area fractions $\phi$ (or densities for short)
are shown in Fig.~\ref{Fig:snapshots}.
At low densities the system is gas-like where no long-lived clusters exist (see also movie M1 in the Supplemental Material \cite{supp}).
In contrast, at very high densities the swimmers aggregate and form a global cluster with hexagonal packing (M2 in \cite{supp}).
At intermediate densities the transition between gas-like and crystalline phase occurs.
Interestingly, the collective structure of the system significantly depends on the hydrodynamic near field
the squirmers create around each other and which is characterized by the squirmer parameter $\beta$.
In particular, for $\beta=0$ and area fractions $\phi \gtrsim 0.5$ hexagonal clusters emerge and the system 
clearly separates in the gas-like and crystalline phase.
In such a phase-separated state a single cluster forms similar to observations in Brownian disks \cite{Fily12,Redner13}.
The cluster is very dynamic since particles leave and join and the cluster re-arranges permanently (M3 in \cite{supp}).
There is a pronounced difference between pushers and pullers.
While pushers move in a more uniform phase (M4 in \cite{supp}) and ultimately develop a single cluster at
$\phi=0.64$ ($\beta = -1.5$) or $\phi = 0.83$ ($\beta = -3$),
pullers rather form several hexagonal structures (M5 in \cite{supp}) and only develop a single cluster at the highest density.

\begin{figure}
\includegraphics[width=\columnwidth]{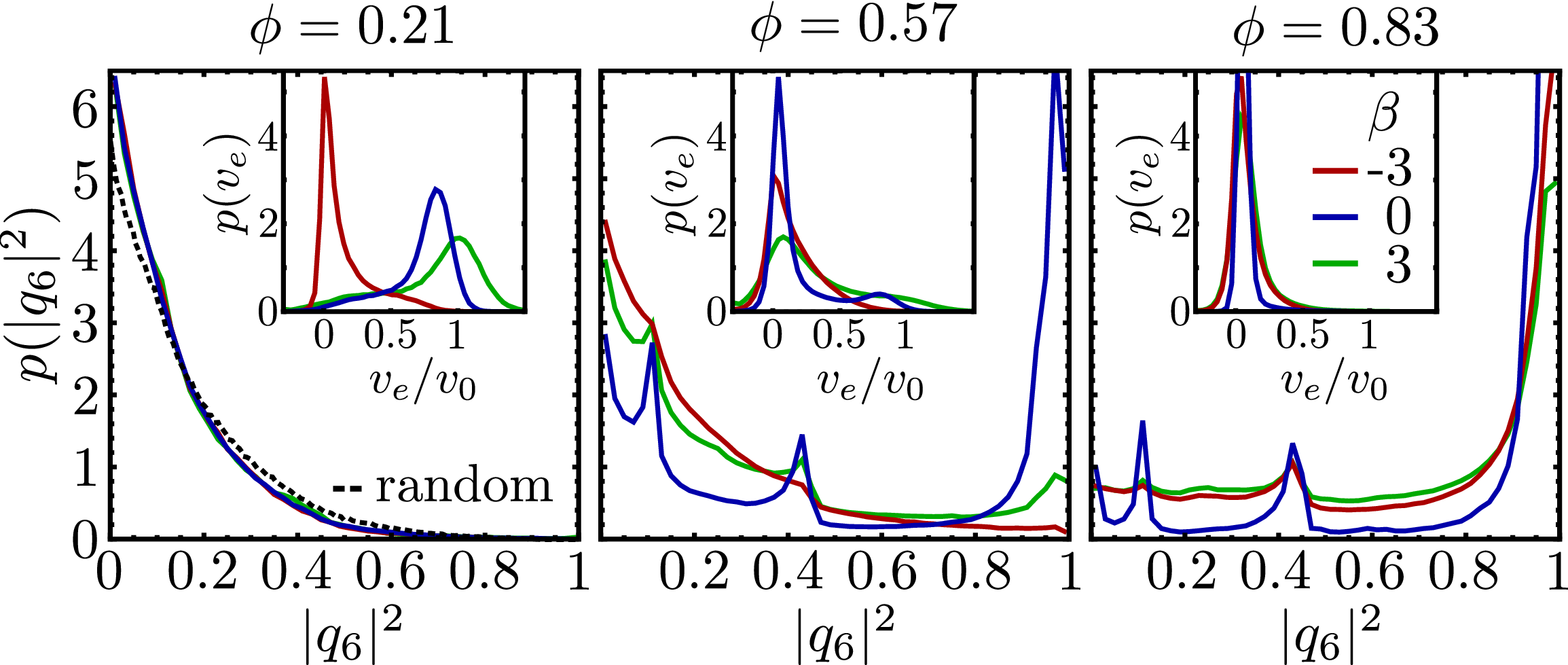} 
\caption{
Distributions of local bond parameter $|q_6|^2$ and swimmer velocities $v_e$ (insets)
at low ($\phi=0.21$), medium ($\phi=0.57$) and high ($\phi=0.83$) density for different
squirmer types: $\beta =0$ (blue), $\beta = -3$ (red), and $\beta = 3$ (green).
}
\label{Fig:H}
\end{figure}

To quantify our findings, we identify hexagonal clusters by introducing the bond parameter $|q_6^{(k)}|^2\in [0,1]$
that measures local 6-fold bond orientational order of particle $k$, where
$ q_6^{(k)} := \frac 1 6 \sum_{j \in N_6^{(k)}} e^{i6\alpha_{kj}}$.
Here the sum goes over the six nearest neighbors of particle $k$,
and $\alpha_{kj}$ is the angle between $\mathbf{r}^{(k)}-\mathbf{r}^{(j)}$ and a randomly chosen axis \cite{Steinhardt83,Bialke12}.
We show the color-coded bond parameter $|q_6|^2$  in the snapshots of Fig.~\ref{Fig:snapshots}.
In Fig.~\ref{Fig:H} we also plot the corresponding distributions $p(|q_6|^2)$ for different $\beta$ and $\phi$.
At low area fraction $\phi=0.21$ in the gas-like state, $p(|q_6|^2)$ is similar to what one expects for randomly distributed particles.
However, at the intermediate density $\phi=0.57$ the additional pronounced 
maximum at $|q_6|^2 \approx 1$ for the neutral squirmer ($ \beta = 0$) shows coexistence of pronounced
hexagonal clustering with the gas-like state and indicates phase separation. Finally, at high density $\phi=0.83$ 
most of the squirmers reside in a hexagonal cluster.
Note that the peaks at $|q_6|^2 \approx 4/9$ and $|q_6|^2 \approx 1/9$ result from particles at the border of this cluster.

\begin{figure}
\includegraphics[width=.65\columnwidth]{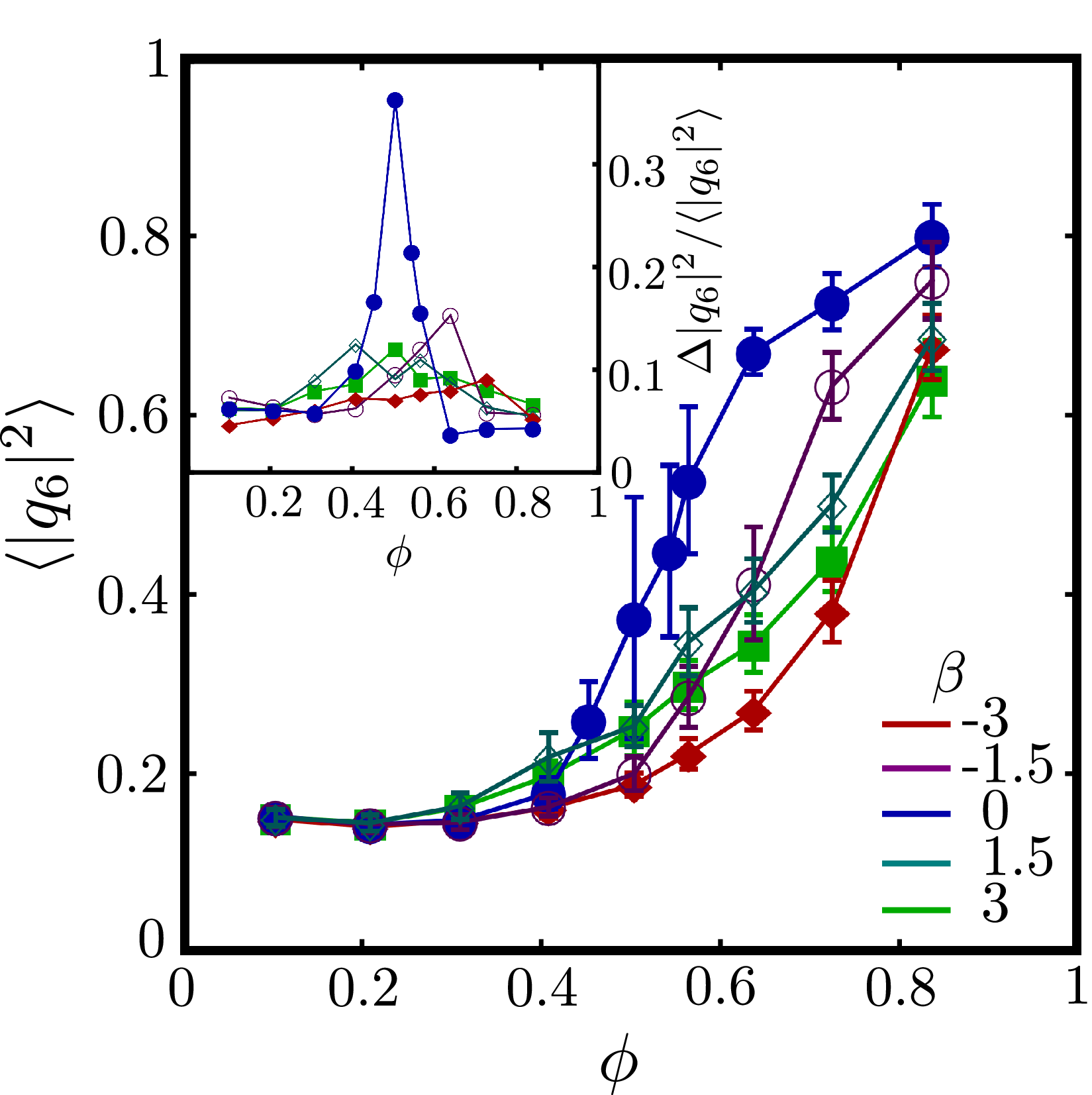} 
\caption{
Bond-orientational
order parameter $\langle |q_6|^2  \rangle$
plotted versus area fraction $\phi$ for pushers ($\beta<0$), pullers ($\beta>0$) and
neutral squirmers ($\beta=0$). Inset:  Order parameter fluctuations $\Delta |q_6|^2/ \langle |q_6|^2  \rangle$.}
\label{Fig:q6}
 \end{figure}

In addition, we introduce the \textit{mean} local bond orientational order $\langle |q_6|^2 \rangle $ as a structural order parameter. 
Figure \ref{Fig:q6} shows $\langle |q_6|^2 \rangle $ plotted versus area fraction $\phi$ for several squirmer parameters $\beta$.
While the curves for $\beta=0$ and $\beta=-1.5$ have a sigmoidal shape where the steep region indicates phase separation, the transition
between the gas-like and crystal phase occurs more smoothly for other $\beta$.
The inset of Fig.~\ref{Fig:q6} shows the fluctuations of the bond-orientational order parameter.
The pronounced maximum for $\beta=0$ at the transition point $\phi \approx 0.50$
is due to hexagonal clusters which form and decay as  shown in movie M6 \cite{supp}.
Both the sigmoidal shape and the strong fluctuations near the critical density 
indicate the existence of a non-equilibrium phase transition.
However, since we are far away from the thermodynamic limit we are not able to determine the order of this transition.

We also show distributions of the velocity component along the squirmer orientation,
$v_e^{(i)} = \mathbf{v}^{(i)} \cdot \mathbf{e}^{(i)}$ (see insets in Fig.~\ref{Fig:H}), which 
demonstrate the \textit{effective activity} of the swimmers.
As expected, at low density the distributions show a maximum at high $v_e$ except for strong pushers ($\beta=-3$)
since they orient normal to the bounding walls as we will discuss below.
For $\beta=0$ the bimodal form at $\phi=0.57$ confirms the phase-separated state
but already the majority of the squirmers are in the cluster phase.
Finally, at high densities most of the squirmers block each other and hence do not move which is indicated by the 
sharp peaks around $v_e\approx 0$.

In order to illustrate the influence of hydrodynamic interactions between the squirmers,
we compare our results with Brownian dynamics (BD) simulations where hydrodynamic interactions are absent. 
We perform both BD simulations in quasi-2D, where the active spheres are free to rotate in 3D  \cite{supp},
and pure 2D simulations as described in \cite{Redner13}, see  Fig.~\ref{Fig:snapshots}.
As reported previously, active Brownian disks in 2D phase-separate at a sufficiently high density \cite{Fily12,Redner13}
which we also quantified by a sigmoidal shape of $\langle |q_6|^2 \rangle (\phi) $ similar to the $\beta =0$ squirmer
(curve not shown in Fig.~\ref{Fig:q6}).
Nevertheless, the clusters are less stable compared to the squirmers (M7 in \cite{supp}).
In contrast, active Brownian spheres in quasi-2D only develop small and short-lived clusters (M8 in \cite{supp})
and rather behave like pullers.

Literature mentions three conditions that favor phase separation. First, it occurs at sufficiently large swimming 
speeds of the active particles so that they collide frequently. The swimmers form clusters where they become trapped 
since they point towards the cluster center \cite{Redner13,Buttinoni13}. Secondly, rotational diffusion has to be 
sufficiently small such that the particles stay trapped otherwise crystal nucleation is hindered \cite{Redner13,Buttinoni13}. 
Thirdly, a slow-down of the swimming speed also stabilizes the condensed cluster phase \cite{Tailleur08,Cates13}.
We now use this conditions to give some qualitative arguments for the observed subtle phase behavior.

\begin{figure}
\includegraphics[width=0.98\columnwidth]{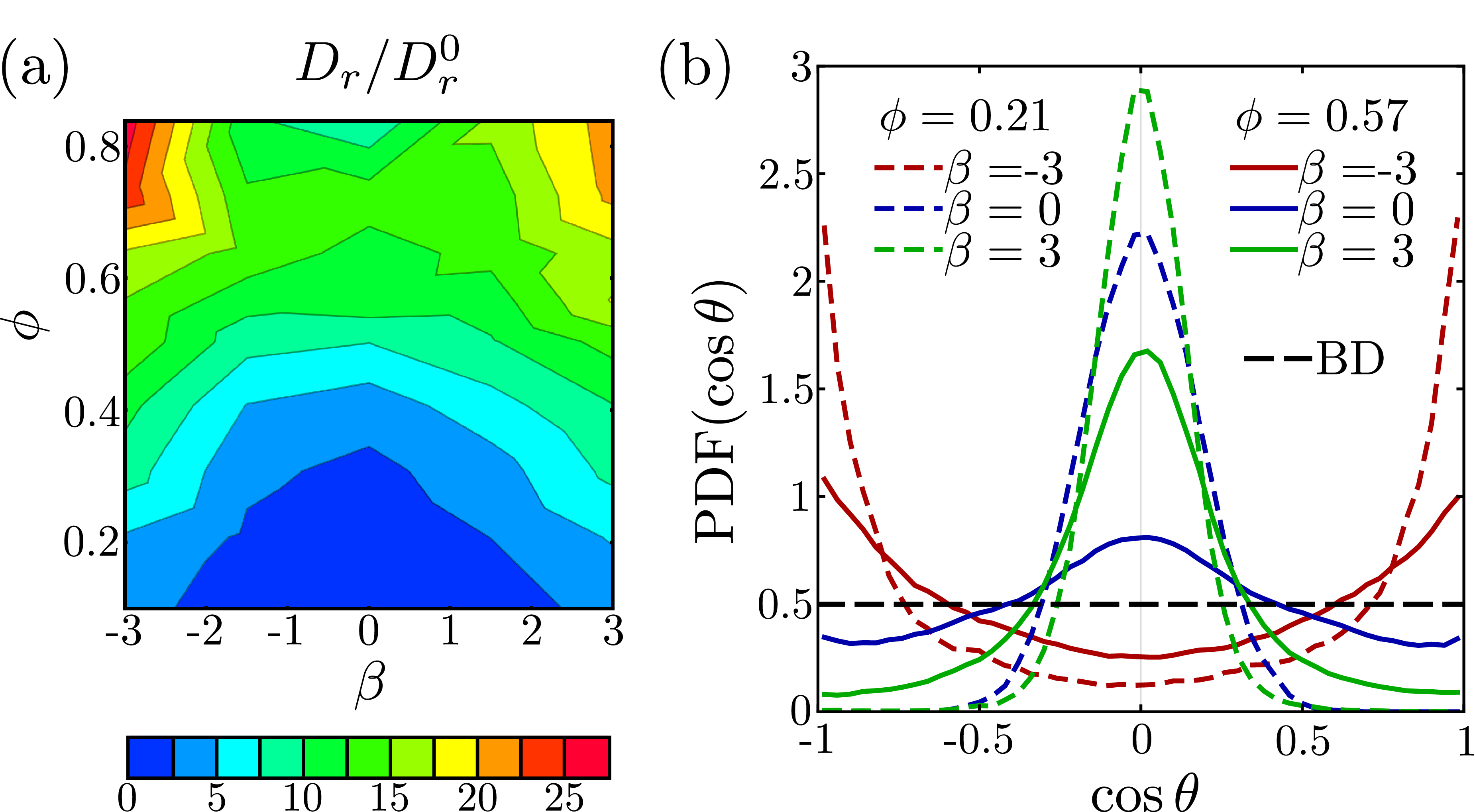}
\caption{
(a) Rotational diffusion coefficient in units of its thermal value $D_r^0$ versus area fraction 
$\phi$ and squirmer parameter $\beta$.
(b) Distributions of squirmer orientations for two area fractions $\phi$. $\theta$ is the angle with respect to
the wall normal.  
}
\label{Fig:CE}
\end{figure}

In our MPCD simulations the reorientation of the microswimmers is mainly determined by the 
hydrodynamic near field between them.
Since this reorientation occurs stochastically, it strongly contributes to rotational diffusion \cite{Ishikawa08a}.
We measure the rotational diffusion coefficient $D_r$ using an exponential fit for the 
orientational correlation function $\langle \mathbf{e}^{(i)}(0)\cdot \mathbf{e}^{(i)}(t)  \rangle = e^{-2D_r t}$
where we average over all swimmers and over all runs.
Depending on $\phi$  and $\beta$ the rotational diffusion coefficient $D_r$ is enhanced compared to its thermal 
value $D_r^0$, in our system up to a factor  $D_r/D_r^0 \approx 25$ for strong pushers at very high densities [Fig.~\ref{Fig:CE}(a)].
Due to lubrication torques acting between two squirmers, the angular velocity of a squirmer in the vicinity of 
the second squirmer depends on the difference of their surface velocities [Eq.~(\ref{Eq:1})] at the points of closest approach \cite{Ishikawa06}.
Lubrication torques increase with area fraction $\phi$ and the averaged magnitude of the surface velocity with $|\beta|$.
This explains the major trends in the rotational diffusivity $D_r$ in Fig.\ \ref{Fig:CE}(a).

Apparently, for small $|\beta|$ and at a bare Peclet number $\mathrm{Pe} = 360$, $D_r$ is sufficiently small for phase separation to occur.
But why do neutral squirmers clearly phase-separate into gas-like and hexagonal structures
and the active Brownian spheres do not.
The reason is that for neutral squirmers the self-trapping mechanism for forming particle clusters is strongly enhanced.
This is due to the combined action of hydrodynamic swimmer-swimmer interactions that slow down the squirmers at close contact
and hydrodynamic swimmer-wall interactions that strongly influence the squirmer orientation.
Figure~\ref{Fig:CE}(b) shows the distribution of the squirmer orientation along the wall normal,
so $\cos\theta\approx 0$ means orientation in the plane and $\cos\theta\approx\pm 1$ perpendicular to the wall.
[see also Fig.~\ref{Fig:geometry}(b)]. 
The distributions clearly depend on the type of swimmer and on density.
While at low density pullers and neutral squirmers orient in plane of the bounding walls, strong pushers tend to orient perpendicular to 
the walls, in contrast to results from hydrodynamic far-field approximations \cite{Berke08}.
This also explains the velocity distribution for $\beta = -3$ at low density in Fig.~\ref{Fig:H}.
At higher densities the permanent interaction with other swimmers broadens the angular distribution for pullers and even stronger for $\beta = 0$.
So, neutral squirmers also orient perpendicular to the wall and thereby reduce the in-plane velocities in a cluster.
This hinders them to escape the cluster such that self-trapping and cluster growth is enhanced.
(This orientational effect does not occur in the 2D squirmer simulations in \cite{Fielding12} and might be one reason for the suppressed phase separation.)
Active Brownian spheres, on the other hand, have a flat $\cos\theta$ distribution independent of density since their orientations are not
influenced by other particles and bounding walls.
In addition, since they do not experience hydrodynamic pressure acting between them, they do not slow down unless they start to overlap.
This suppresses nucleation although their rotational diffusion constant is smaller compared to neutral squirmers.

In Fig.~1 of \cite{supp} we combine both rotational diffusion and mean in-plane orientation by plotting separate
effective persistence numbers $\mathrm{Pe}_r = v_0 \langle \sin \theta \rangle /(2R D_r)$ versus $\phi$ for the gas-like and the cluster phase.
In particular, for the neutral squirmer a large $\mathrm{Pe}_r(\mathrm{gas})$ and a sharp drop towards $\mathrm{Pe}_r(\mathrm{cluster})$ at $\phi = 0.5$ 
clearly indicates the onset of phase separation.

We can calculate the preferred squirmer orientations at low densities [Fig.~\ref{Fig:CE}(b)] using the lubrication approximation for
a single squirmer confined between two parallel walls \cite{supp}.
The analysis confirms that a puller or neutral squirmer move, respectively, stable or marginally stable parallel to the walls while
a sufficiently strong pusher has a stable orientation perpendicular to the walls \cite{Zhu13b}.
Thermal motion and squirmer interactions then result in the distributions of Fig.~\ref{Fig:CE}(b).
Indeed, it has been demonstrated that thermal noise plays an important role in swimmer-swimmer interactions
even at large persistence numbers \cite{Goetze10,Drescher11}.
Noise might also be the reason why we do not observe polar order while deterministic squirmer simulations in bulk show it
\cite{Ishikawa08a,Evans11,Alarcon13}.

To conclude, using the squirmer as a model swimmer whose type can be tuned by the stresslet parameter $\beta$, 
we have shown that hydrodynamic near fields determine the phase behavior of active particles in an 
experimentally relevant quasi-2D geometry. These near fields cause a pronounced increase of rotational diffusion,
slow down squirmers during collision, and influence the squirmer orientation which can enhance the 
self-trapping in crystalline clusters.
Neutral squirmers phase-separate in a gas-like and cluster phase accompanied by a strong decrease in motional persistence in the cluster phase.
In contrast, strong pushers and pullers gradually develop the hexagonal cluster.
Our approach can be extended to study the collective motion of active Janus particles which have a different
slip-velocity profile at their surface \cite{Janus}.

We thank
Francisco Alarcon,
Philipp K\"{a}hlitz,
Alexander Morozov,
Sriram Ramaswamy,
Sebastian Reddig,
Konstantin Schaar,
Paul van der Schoot,
Shashi Thutupalli,
John Toner,
Katrin Wolff,
and
Julia Yeomans
for helpful discussions.
We acknowledge funding from the DFG within the research training group GRK 1558 and 
by grant STA 352/10-1.


\end{document}